\documentclass[12pt]{article}
\usepackage{graphicx}
\usepackage{amsmath}
\usepackage{authblk}
\usepackage[round]{natbib}

\title{Predicting patterns of long-term adaptation and extinction with population genetics}

\author[1]{J Bertram}
\affil[1]{Department of Ecology and Evolutionary Biology, University of Arizona, Tucson, AZ 85721}
\author{K Gomez}
\affil[2]{Program in Applied Mathematics, University of Arizona, Tucson, AZ 85721}
\author[1]{J Masel}
\begin{document}

\maketitle

To whom correspondence should be addressed. E-mail: jbertram@email.arizona.edu and masel@email.arizona.edu.

Running title: Population genetics of gradual extinction


\pagebreak


\begin{abstract}
Population genetics struggles to model extinction; standard models track the relative rather than absolute fitness of genotypes, while the exceptions describe only the short-term transition from imminent doom to evolutionary rescue. But extinction can result from failure to adapt not only to catastrophes, but also to a backlog of environmental challenges. We model long-term evolution to long series of small challenges, where fitter populations reach higher population sizes. The population's long-term fitness dynamic is well approximated by a simple stochastic Markov chain model. Long-term persistence occurs when the rate of adaptation exceeds the rate of environmental deterioration for some genotypes. Long-term persistence times are consistent with typical fossil species persistence times of several million years. Immediately preceding extinction, fitness declines rapidly, appearing as though a catastrophe disrupted a stably established population, even though gradual evolutionary processes are responsible. New populations go through an establishment phase where, despite being demographically viable, their extinction risk is elevated. Should the population survive long enough, extinction risk later becomes constant over time. 
\end{abstract}

Extinction has historically been viewed in two different ways \citep{smith_89,raup_1994,macleod_2014}: the ``catastrophic'' view, which revolves around sudden, severe disturbances; and the ``gradualist'' view, which emphasizes long-term evolutionary processes such as failure to adapt to slowly deteriorating circumstances. While catastrophes are bound to to occur eventually, and present an obvious danger when they do, the threat posed by cumulative changes in the environment (both biotic and abiotic) is no less serious. Although the deleterious effects of these changes can be partly mitigated by physiological or behavioral adaptation, if they are not offset by evolutionary adaptation, and begin to accumulate, extinction is inevitable \citep{burger_95}. 

The catastrophic and gradualist views are not mutually exclusive. A population's vulnerability to additional disturbances depends on its current burden of adaptive failures or ``lag load'' (a measure of the fitness distance between a genotype and a perfectly adapted genotype; \citealt{smith_76}), which may have accumulated gradually. Thus, even in cases where extinction is proximately caused by major disturbances, long-term evolution may have exerted a strong influence on the extinction process. Yet relatively little is known about gradual extinction processes \citep{burger_95}, especially when compared to the celebrity of catastrophes in paleontology \citep{raup_1994,macleod_2014} and conservation biology \citep{barnosky_2011}. 

We identify three main existing classes of gradual extinction model (Table 1). First, applying classical population genetics theory, \cite{haldane_57} gave the first quantitative theoretical predictions of the time scales and risks associated with long-term evolution by considering the number of deaths attributable to selection during a single selective substitution --- the ``cost'' of selection. Measured as a proportion of population size ($N$), this gives an estimate of the fitness reduction during substitution (``substitutional load''), or the number of generations needed for substitution. While not directly predicting extinction risk, substitutional load arguments attempt to identify limits to the rate of adaptation. 

Second, probably the most well-known model of gradual extinction is B\"{u}rger and Lynch's quantitative genetics model of stabilizing selection on a single trait, where environmental change is represented by change in the optimal trait value \citep{burger_95,gomulkiewicz_2009}. Population size is finite, so extinction will occur eventually because of demographic stochasticity, regardless of environmental change. However, extinction occurs much more rapidly when the environmental change rate exceeds a critical value at which the mean phenotype lags so far behind its optimum that demographic decline ensues (mean absolute fitness falls below one). \cite{burger_95} calculated this critical environmental change rate as well as times to extinction, although individual-based simulations were needed for most of their predictions.

Third, the adaptive dynamics approach has been used to explore the consequences of feedbacks between evolution and ecology in communities of evolving species  \citep{dieckmann_2004}. Adaptive dynamics describes evolution as a sequence of ``trait substitutions'', in which one species at a time in a community is invaded by an adaptive mutant (each species has a single trait value), moving the community from one ecological equilibrium to a neighbouring one. The consequences for extinction can be dramatic; species may drive themselves extinct via trait substitution sequences (``evolutionary suicide''), even in the absence of abiotic environmental change or evolutionary change in other species \citep{ferriere_2012}. The published extinction predictions of adaptive dynamics have so far been primarily descriptive \citep{ferriere_2012}. 

Here we present a new model of long-term adaptation and extinction that builds upon and extends previous work. Like \cite{burger_95}, we assume that extinction is driven by gradual environmental deterioration. Extinction can be avoided by evolutionary adaptation, which depends on genetic and demographic factors. However, our model is based on population genetics rather than quantitative genetics, and is not restricted to quantitative traits. As a result, individual mutations can have large or intermediate effects in our model rather than only modifying quantitative trait loci of small effect; the former are known to be important drivers of adaptation \citep{orr_2005}.

Similar to adaptive dynamics, we recognize the importance of feedbacks between long-term evolutionary changes and the short-term demographic response of the population. We restrict our attention to changes in the focal population size $N$ without modeling the complex response of entire ecological communities. Poorly adapted populations will generally have fewer individuals, which reduces adaptive mutant production and increases the chance of further fitness decline, reminiscent of a ``mutational meltdown'' \citep{lynch_93}. However, in low fitness populations, more beneficial mutations will be available (there are more problems to be addressed). Each beneficial mutation will also have a greater effect compared to when fitness is high (diminishing returns epistasis; \citealt{wiser_2013}). 

Our model is based on Desai and Fisher's (2007) asexual traveling wave
model, under which the steady state adaptation rate is determined by a bal-
ance between selection and beneficial mutations. In that model, as in most
population genetic models, $N$ is constant over time and evolution occurs
along a relative fitness axis; thus extinction is impossible. To model extinction, we replace relative fitness with a simple model of density-dependent
absolute fitness, so that $N$ changes dynamically. The beneficial mutation rate is assumed to depend on absolute fitness. In addition, we introduce a Markov chain approximation for the population's long-term evolution, which is considerably simpler than the full traveling wave model.

We use our model to explore a few basic questions about long-term evolution: (1) What are the conditions for long-term persistence, and are persistence times predicted from our micro-evolutionary model consistent with macro-evolutionary persistence times in nature? (2) What is the distribution of extinction times? (3) Should we expect to be able to distinguish gradual from catastrophic causes of extinction based on observations of a population's behavior prior to extinction?

\renewcommand{\figurename}{Table}
\begin{figure}
\centering
\label{fig:table}
\includegraphics[scale=0.6]{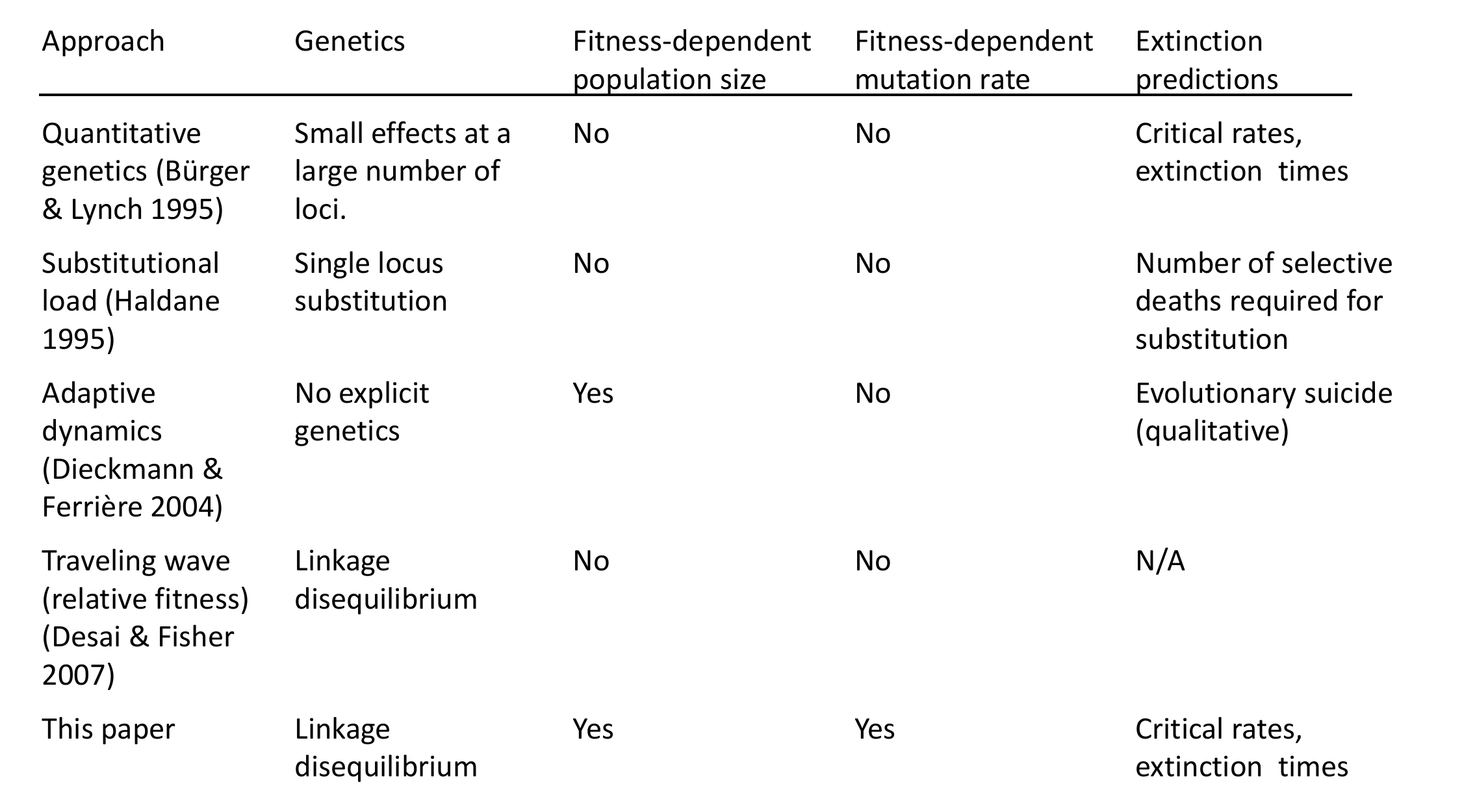}
\caption{Comparison of this paper with related models.} 
\end{figure}

\section*{Model}

Following Desai and Fisher (\citeyear{desai_2007}), the population is divided into discrete fitness classes differing by multiples of a constant fitness increment $s$ (Fig.\ \ref{fig:travellingwave}). Population size $N$ is assumed to be large enough that the abundances of most fitness classes behave deterministically. Desai and Fisher (\citeyear{desai_2007}) is formulated in terms of relative fitness, and $N$ is constant. We instead use a simple logistic model of absolute fitness,
\begin{equation}
\frac{1}{n_i}\frac{d n_i}{d t}=b\left(1-\frac{N}{\kappa}\right) - (d+is). \label{eq:dni_bd}
\end{equation}
Here $n_i$ is the abundance of fitness class $i$ (so that $N=\sum_i n_i$), $b$ is the intrinsic birth rate, $d$ is the mortality rate of perfectly adapted ($i=0$) individuals, and $i s$ is the additional mortality associated with fitness class $i$. The indices $i$ count the number of fitness classes from perfection, and can therefore be interpreted as a measure of lag load \citep{smith_76}. $\kappa$ is the maximum possible population size without deaths, representing territorial or resource limitations. Henceforth, the index $i$ will be used to refer to any possible fitness class, which could be empty, while $j$ will be used to refer specifically to the most fit class that has abundance large enough that it behaves approximately deterministically according to Eq.\ \eqref{eq:dni_bd} (Fig.\ \ref{fig:travellingwave}).

$\kappa$ is distinct from the maximum achievable abundance (the carrying capacity), where births balance deaths ($d N/d t=0$). If all individuals are in the same fitness class $j$, the carrying capacity is $K_j=(b-d-j s) \kappa/b$, which decreases with decreasing fitness. The population is not viable if deaths exceed births at low population density i.e. $d+i s>b$ for all occupied fitness classes. This defines an extinction threshold $i_e\approx (b-d)/s$, given by the $i$ where $d+i s$ first exceeds $b$.  

We only consider beneficial mutations, and all mutations have the same fitness effect $s$ irrespective of genetic background. We assume that the beneficial mutation rate in fitness class $i$ is $Ui$ per birth, where $U$ is a constant. This represents a ``running out of mutations'' effect, where there are more ways for genetic novelty to improve fitness in poorly-adapted genotypes (and no ways to improve a perfect genotype). We return to our running out of mutations assumption, specifically how it differs from diminishing returns epistasis, in the Discussion.

\setcounter{figure}{0}
\renewcommand{\figurename}{Fig.}
\begin{figure}[h]
\includegraphics[scale=0.6]{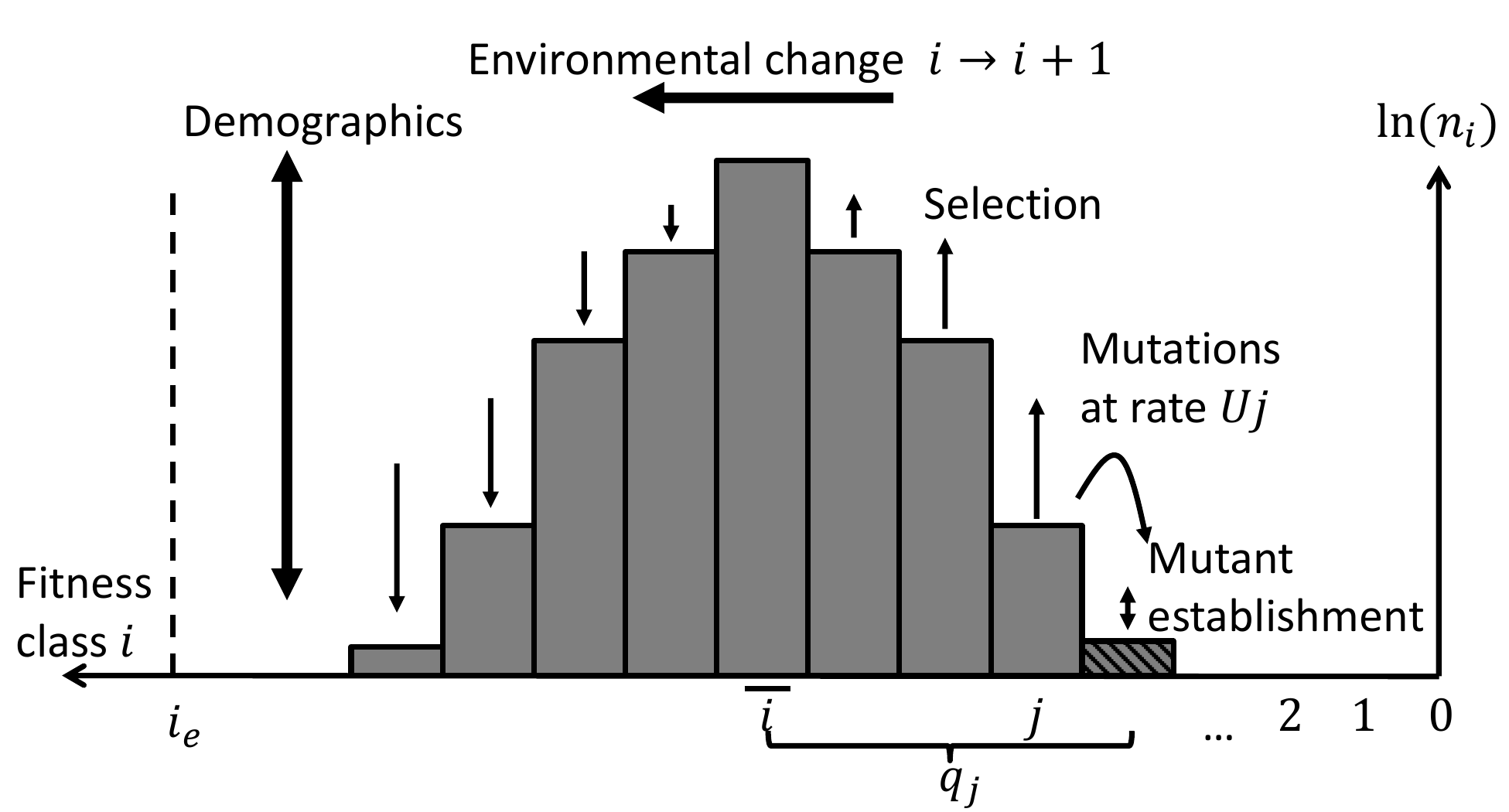} 
\caption{\label{fig:travellingwave} Absolute fitness classes $i$ representing fitness increments of size $s$, with abundances $n_i$. Environmental deterioration intermittently reduces population fitness by $s$. Fitness classes grow or decline relative to each other depending on whether their fitness is respectively greater or smaller than the mean fitness $\overline{i}$ (small vertical arrows). Population size $N$ changes dynamically with fitness (double-headed vertical arrow). At the nose of the distribution, mutant establishment is stochastic (hatched bar). The fittest established class is $j$, and mutants are $q_j=\overline{i}-(j-1)$ fitness classes away from the mean (their fitness advantage is $q_js$). The extinction threshold $i_e$ is shown with a vertical dashed line.}
\end{figure}

There is no sex, so mutations only matter in the leading deterministic class $j$, producing mutants appear in the stochastic ``nose'' $j-1$. Mutations on poorer genetic backgrounds --- away from the nose ---  are doomed to be outcompeted by nose mutants (multiple-mutations interference; \citealt{desai_2007}). Thus, the only relevant mutation rate in our model is that feeding the nose $Uj$. 

Mutant lineages initially have low abundance (starting from a solitary mutant), and are therefore strongly affected by demographic stochasticity. Only some mutant lineages avoid going extinct in the initial stochastic phase and attain a large enough abundance that they grow deterministically  according to Eq.\ \eqref{eq:dni_bd} (a process called ``establishment''). The probability of establishment at the nose, denoted $p_{j-1}$, is approximately $q_j s/(d+js)$ for most mutations (Appendix A), where $q_j=\overline{i}-(j-1)$ is the number of fitness classes that the nose is ahead of the mean (Fig.\ \ref{fig:travellingwave}). However, $p_{j-1}$ can be substantially smaller if environmental change occurs during the establishment process. The calculation of $p_{j-1}$ in this case is discussed in Appendix A.

Once a mutant lineage established, it becomes the new most fit established class  with dynamics governed by Eq.\ \eqref{eq:dni_bd}. The initial abundance for deterministic growth $\nu$, which is applied at the time that the mutation occurs, is a random variable that represents the stochasticity in the time that the mutant lineage takes to establish. The cumulative density function for $\nu$ is given by \citep[Eq.\ 40]{uecker_2011}
\begin{equation}
P(\nu\leq\nu_0)=1-e^{-\nu_0 p_{j-1}},\label{eq:nu_dist}
\end{equation}
so that the mean of $\nu$ is $1/p_{j-1}$ \cite[also see][Eq. 16]{desai_2007}. 

Note that there is a clean separation between the deterministic bulk obeying Eq.\ \eqref{eq:dni_bd}, with fittest class $j$, and the stochastic nose in fitness class $j-1$, as shown in Fig. \ref{fig:travellingwave}. This clean separation holds when establishing mutants make up a small fraction of the population ($Ns\gg 1$), and $Uj$ is small enough that mutant lineages rarely produce double-mutants before establishment (${\rm birth\,rate}\times Uj\ll s$; \citealt{desai_2007}), as is the case here.

Environmental deterioration occurs in discrete events where the entire fitness distribution is shifted backwards by one fitness class ($i\rightarrow i+1$ for all of the population's fitness classes). These events are assumed to follow a Poisson process with mean time $T$ between successive changes. Thus, in addition to being a measure of lag load, $i$ can also be interpreted as the number of environmental challenges facing the individuals in fitness class $i$. In this interpretation, the linear dependence of the mutation rate $Uj$ on $j$ can be interpreted as saying that each environmental deterioration event opens up one new possible beneficial mutation that addresses the new environmental challenge.

Environmental challenges may be biotic or abiotic in our model; fitness differences simply represent differences in mortality without specifying causes. We could easily attribute fitness differences in Eq. \eqref{eq:dni_bd} to births or a mixture of births and deaths instead, but this would not substantially alter our model's behavior.

\section*{Results}

The model described above is simulated numerically (implementation is summarized in Supplement A). In addition to these simulations, we show that the population's long-term evolution can be approximated with a much simpler discrete-time Markov chain (MC).

\subsection*{Markov chain approximation}

\begin{figure}
\includegraphics[scale=.6]{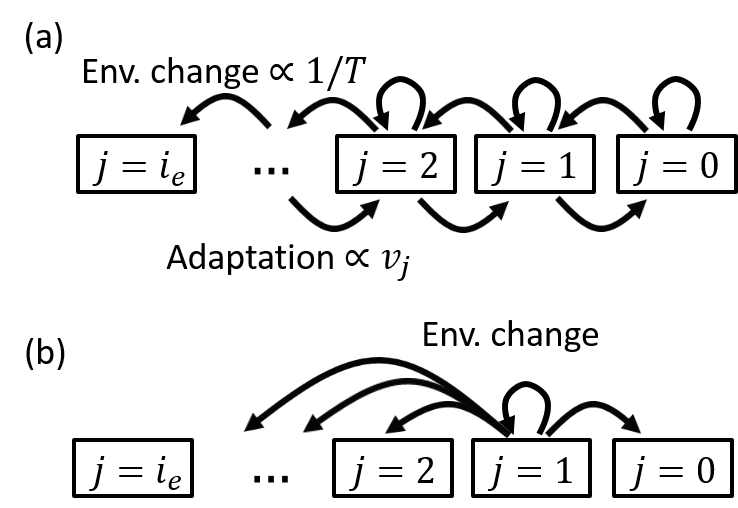} 
\caption{\label{fig:chainfigure} In addition to simulations of the traveling wave model illustrated in Fig.\ \ref{fig:travellingwave}, two Markov chain approximations are used to model long-term evolution. (a) Memoryless adaptation with $j\rightarrow j-1$ transition probability proportional to $v_j$ where $j$ is the fittest established class, and $v_j$ is given by Eq.\ \eqref{eq:vsuccessional} (successional regime) and Eq.\ \eqref{eq:vmm} (multiple-mutations regime). (b) Periodic adaptation in the multiple mutations regime, where each iteration represents the establishment timescale $1/v_j$. Exactly one adaptation event occurs each iteration, but a variable number of environmental change events can occur.}
\end{figure}

Our model has two distinct adaptive regimes: the ``successional'' regime, and the ``multiple mutations'' regime. We first describe our MC approximation in the ``successional'' regime, where fixation (the growth of a newly established mutant to a frequency of $1$) is much faster than the typical time between mutant establishments. The population spends most of the time in equilibrium with all individuals in one fitness class $j$ ($N\approx K_j$), waiting for adaptive mutant establishment or environmental change. Adaptive advances occur at a mean rate $v_j$ equal to the equilibrium birth rate $K_j b(1-K_j/\kappa)$ multiplied by the mutation rate $Uj$ and establishment probability $p_{j-1}$,
\begin{equation}
v_j=K_j b(1-K_j/\kappa) Uj p_{j-1}. \label{eq:vsuccessional}
\end{equation}
Our MC approximation amounts to taking regular ``snapshots'' at intervals given by the characteristic fixation time (Appendix B). In the vast majority of snapshots, the population will be in equilibrium, and will jump between fitness classes with per-snapshot probabilities proportional to $1/T$ for environmental change $j\rightarrow j+1$, and $v_j$ for adaptive mutant fixation $j\rightarrow j-1$. These are the MC transition probabilities. The current MC state is $j$, with possible values $i=0,\ldots,i_e$ (Fig.\ \ref{fig:chainfigure}a), and each MC iteration represents the fixation time. The extinction threshold $i_{e}$ is an absorbing state because the corresponding $N=0$ equilibrium is attained within a single iteration (the scenario where the extinction threshold has been crossed but the population manages to recover by producing higher-fitness individuals with $i<i_e$, called ``evolutionary rescue'', is not possible in our MC approximation).

In the ``multiple mutations'' regime, fixation is slower than the rate at which mutants destined for establishment are produced, and there is standing fitness variation ($q_j>0$). By invoking beneficial mutation-selection balance, the steady-state adaptation rate can be approximated analytically for a given mutation rate, population size and establishment probability \citep{desai_2007}. The latter quantities depend on fitness, particularly the position of the most fit established class $j$. Thus, the corresponding steady state also depends on $j$, giving (Appendix C)
\begin{equation}
q_{j}\approx \frac{2\ln (K_j s)}{\ln{(s/Uj)}},\quad v_j\approx \frac{2\ln (K_j s)-\ln(s/Uj)}{\ln^2(s/Uj)}s. \label{eq:vmm}
\end{equation}
This is a straightforward generalization of Eqs. (40) and (41) in \cite{desai_2007}, which assumed constant mutation rate, population size and establishment probability. 

Unlike the successional regime MC approximation, adaptation cannot be treated as memoryless (independent of the population's history) in the multiple mutations MC approximation. The mutant-generating class $j$ grows over time, and thus so does the overall rate of mutant production at the nose. Consequently, mutant establishment is much less likely shortly after the previous establishment, while class $n_j$ is still small. Moreover, previous growth at the nose is not ``forgotten'' when environmental changes occur. Thus, mutant establishments occur more regularly than memoryless events with mean rate $v_j$. Accordingly, we use two MC approximations to bound the actual behavior of the multiple mutations regime. In the first, we ignore memory so that $v_j$ from Eq.\ \eqref{eq:vmm} is the $j\rightarrow j-1$ transition probability, analogous to the successional case (Fig.\ 2a). In the second, we assume that mutant establishment occurs periodically at given intervals (Fig.\ 2b). Each iteration of the periodic-adaptation MC represents the time required for mutant establishment $1/v_j$, and exactly one establishment happens every iteration. Note that the memoryless and periodic-adaptation MC chains differ only in whether or not \textit{adaptation} is memoryless: in both cases, the transition probabilities are memoryless, as they must be in a Markov chain. The mathematical details of our MC approximation are given in Appendix B.

\begin{figure}
\centering
\includegraphics[scale=0.5]{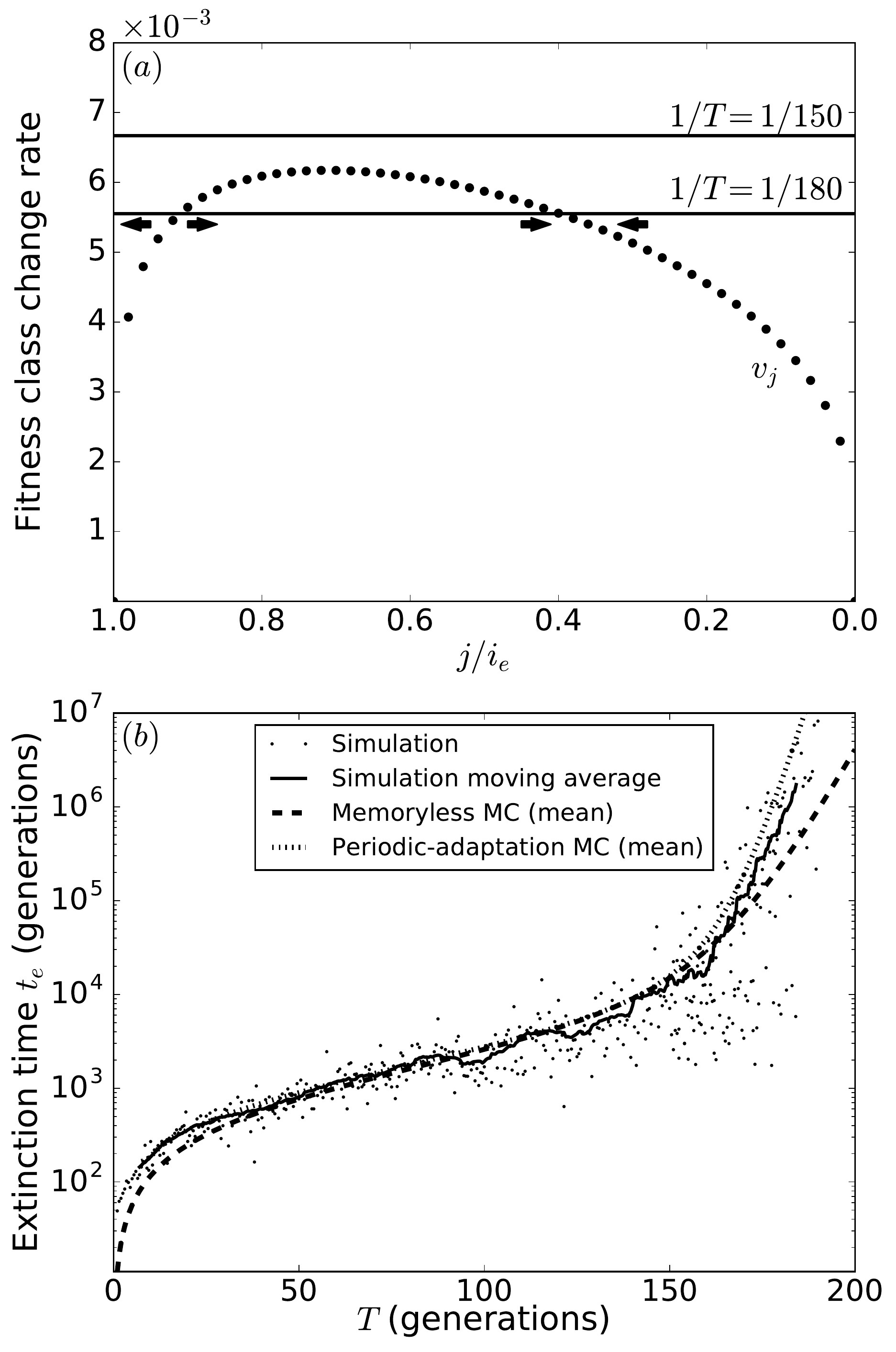} 
\caption{\label{fig:meantimesmm} Time to extinction $t_e$ increases abruptly with increasing $T$. (a) As $T$ increases, the population transitions from ``always losing'' to ``sometimes winning''. Arrows show mean direction of fitness change near points where $1/T=v_j$; $j/i_e\approx 0.4$ is an ``attractor''. (b) Comparison of simulated $t_e$ and mean $t_e$ predicted from MC approximation (Appendix B). Parameters: $b=2,d=1,U=10^{-6},s=0.02,\kappa=4\times 10^6, j_{\rm initial}/i_e=0.4$ (multiple mutations regime).}
\end{figure}

\subsection*{Extinction times}

Long-term evolution is primarily controlled by the difference between the opposing rates of environmental change $1/T$ and adaptation $v_j$, where $v_j$ tends to zero at perfection $j=0$ (no beneficial mutations) and extinction $j=i_e$ ($N=0$), and exhibits a peak between these extremes (Fig.\ \ref{fig:meantimesmm}a). Figure \ref{fig:meantimesmm}b shows the predicted pattern of time to extinction $t_e$ versus $T$ (time is measured in generations, implemented by setting  $d=1$). When environmental change is relatively rapid ($T<150$), the population cannot keep up with environmental deterioration ($1/T>v_j$ for all $j$), and extinction occurs rapidly (thousands of generations). Modestly slowing environmental change ($T=180$) allows the population to beat the environment ($1/T<v_j$) over part of the fitness domain, dramatically increasing the mean and variance in $t_e$ \cite[compare][Fig. 1b]{burger_95}. This sudden transition to long-term persistence occurs at $T\approx 160$. Fig.\ \ref{fig:meantimesmm} is in the multiple mutations regime; the successional regime gives essentially identical results, except with lower persistence times for given $T$ (since, by definition, there are far fewer adaptive mutants establishing).

MC mean extinction times $\langle t_e\rangle$ (Appendix B) closely follow the full simulation results (Supplement A) in Fig.\ \ref{fig:meantimesmm}b. The periodic-adaptation MC performs better than the memoryless MC, confirming the importance of mutant establishment ``memory'' in the multiple mutations regime. 

\begin{figure}
\centering
\includegraphics[scale=0.6]{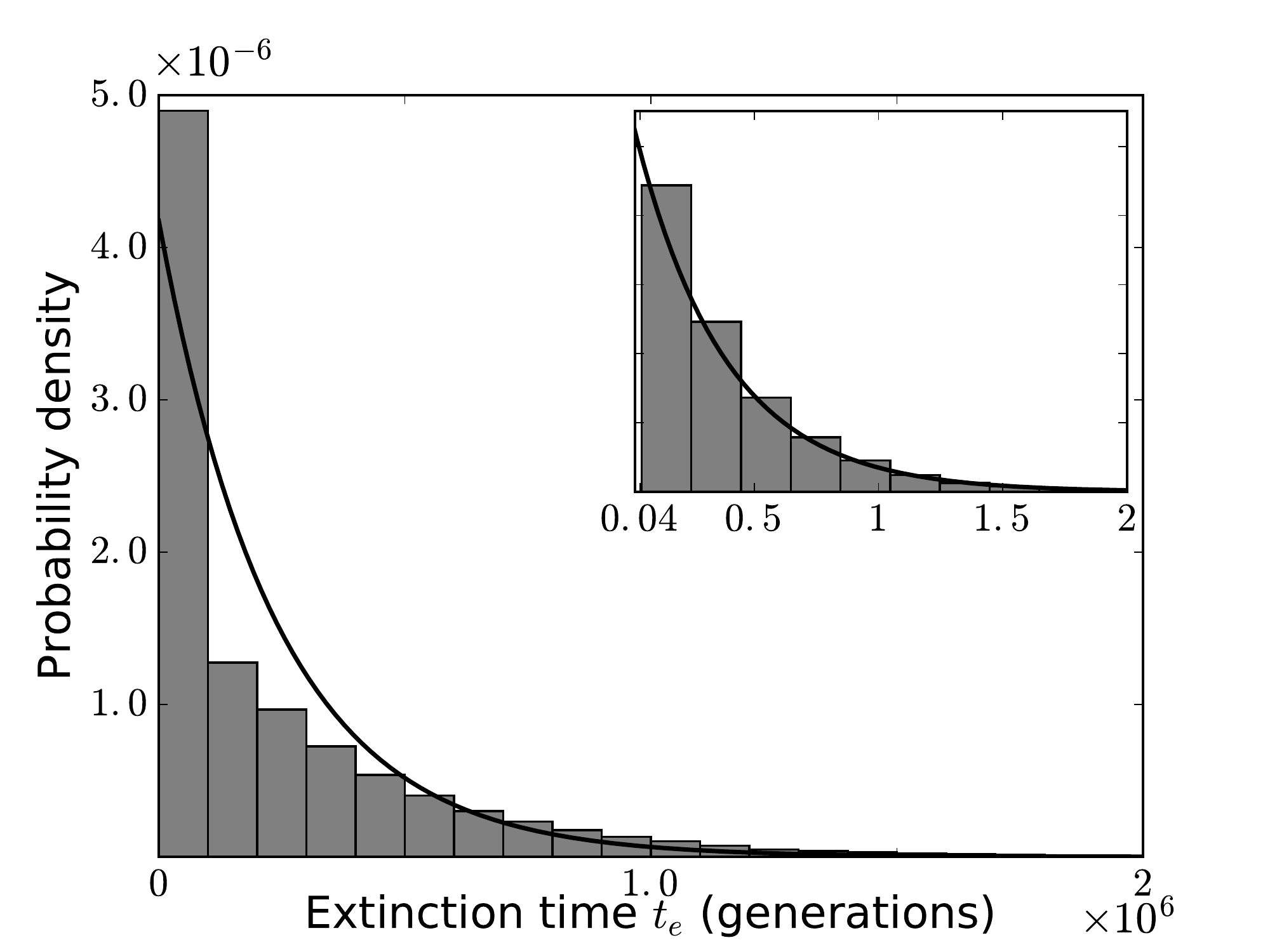} 
\caption{\label{fig:extimehist} The distribution of extinction times for fixed $T$ is not exponential, but has an exponential tail. Main figure: $t_e$ from memoryless MC simulations (histogram; same parameters as Fig.\ \ref{fig:meantimesmm} except $j_{\rm initial}/i_e=0.8$) compared with the corresponding (same mean) exponential distribution (curve). Inset: same simulations omitting the initial spike of rapid extinction $t_e<4\times10^5$ (histogram). Curve shows exponential distribution with mean given by MC mean $t_e$ (using Eq.\ \eqref{eq:meante_linsys}) assuming $j_{\rm initial}/j_e=0.4$ (the attractor in Fig.\ \ref{fig:meantimesmm}).}
\end{figure}

Fig.\ \ref{fig:extimehist}a shows the distribution of extinction times for a population with low initial fitness $j_{\rm initial}/i_e=0.8$ \cite[compare][Fig. 4]{burger_95}. This could represent a newly establishing population at the start of peripatric speciation, for example. As expected, the distribution is sharply peaked near zero, reflecting a high risk of early extinction. The distribution also has a long tail (Fig.\ \ref{fig:extimehist}a inset), reflecting cases where the population manages to reach the stable ``attractor'' at $j/i_e\approx 0.4$ (Fig.\ \ref{fig:meantimesmm}a). Once the attractor is reached, long-term persistence is possible. 

\begin{figure}[h]
\centering
\includegraphics[scale=0.6]{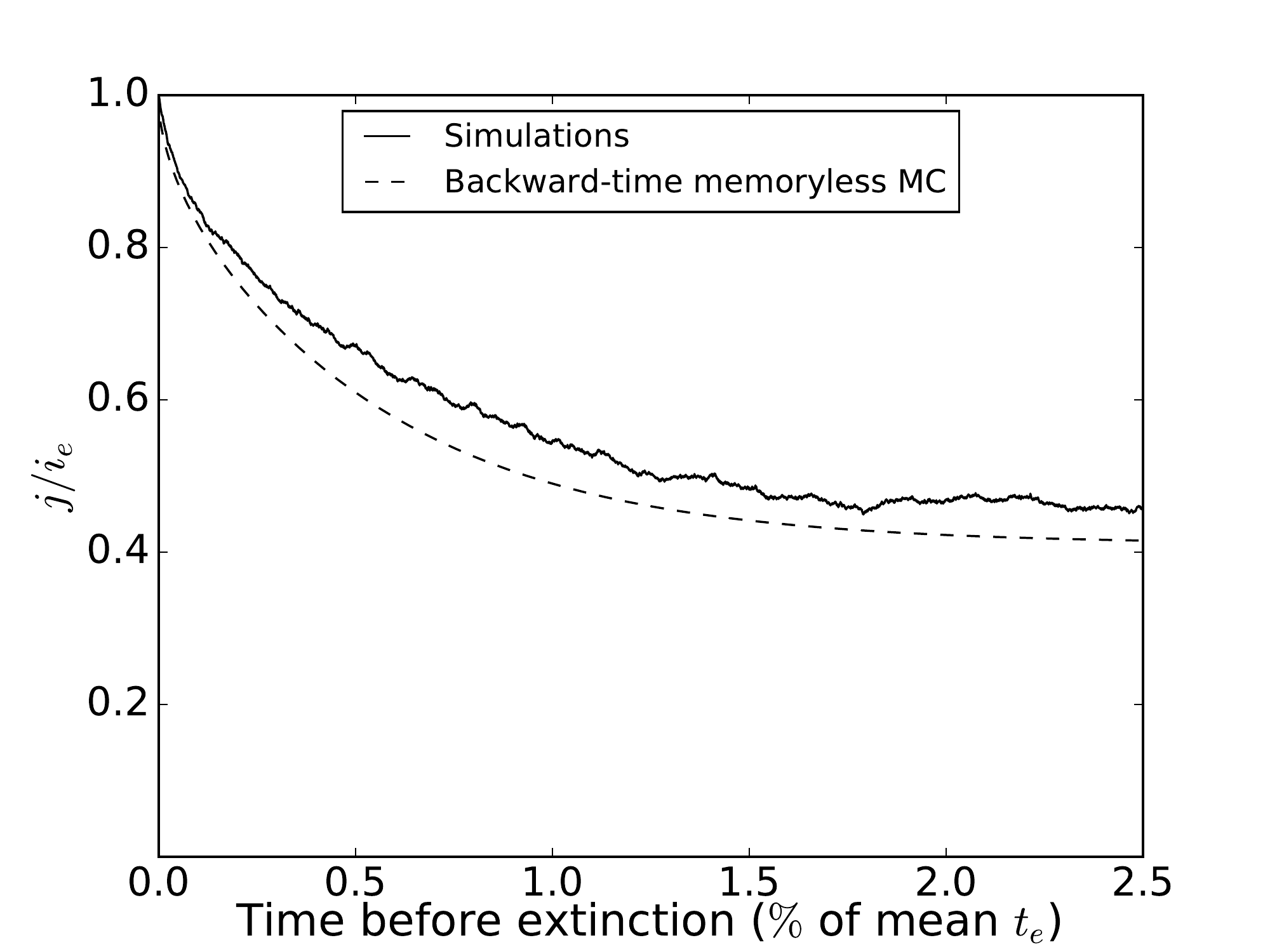} 
\caption{\label{fig:beforeext} For persistent populations, fitness declines rapidly at the time of extinction compared to mean $t_e$. Shown is the expectation of $j$ for the backward-time trajectory immediately preceding extinction, using both backward-time memoryless MC (Supplement B) and averaging over simulated trajectories in the full model. Mean persistence time is $\approx 2\times 10^6$ generations. Same parameters as Fig.\ \ref{fig:meantimesmm} with $T=180$.}
\end{figure}

The tail in the distribution of $t_e$ is exponential (Fig.\ \ref{fig:extimehist}a inset), indicating that extinction risk is effectively constant over time. The reason for this constant risk is that the population remains near the attractor for most of its existence. When extinction does occur, it is due to an abnormally rapid sequence of environmental changes and/or slow sequence of mutant establishments rather than a gradual erosion of fitness. The resulting decline in fitness is rapid compared to the mean persistence time. Fig.\ \ref{fig:beforeext} shows the average decline in fitness immediately prior to extinction obtained by averaging over simulated trajectories in the full model as well as a backward-time variant of our MC approximation (Supplement B); moving from the attractor to the extinction threshold only takes around $1\%$ of mean persistence time. 

This explains why such a substantial discrepancy exists between the memoryless MC and simulations after the transition to persistence ($T>160$), but not before it ($T\approx 100$): after the transition, large fluctuations in the number of adaptive establishments are much more likely when establishments are memoryless. Before the transition, there is no attractor, mean time to extinction is determined by the average decline in fitness due to the fact that $1/T<v_j$ (specifically, $\langle t_e \rangle = j_{\rm initial}/(1/T-v_j)$), and fluctuations are only of secondary importance.

\section*{Discussion}

It is difficult to directly compare our predictions with fossil data because we have only considered a single population adapting to local environmental changes. Fossil extinction times also reflect larger scale processes in which environmental heterogeneity, range shifts and migration are potentially important. Nevertheless, population-level processes should have a strong influence at larger spatial and temporal scales, particularly for species with relatively small ranges. 

For concreteness, consider the example of mollusc species, which feature prominently in the fossil record, and can have tiny geographic ranges \citep{stanley_1986}. A representative generation time for fossil mollusc species is on the order of $10$ years \citep{powell_2011}. With this generation time, our extinction time predictions (Fig.\ \ref{fig:meantimesmm}) are easily large enough to be consistent with typical fossil species persistence times of several million years \citep{raup_1994}. The shortest interval between environmental changes consistent with long-term persistence is $T\approx160$ generations, or approximately a $2\times 10^3$ years between $2\%$ increases in mortality rate (since $d=1$ and $s=0.02$). By comparison, the major glacial cycles in the last $1$ million years occurred at intervals of roughly $10^5$ years \citep{augustin_2004}. Thus, in the absence of adaptation, the mortality rate would roughly double over the course of each major glaciation cycle. This rate of environmental change is certainly gradual compared to catastrophes such as the aftermath of bolide impacts, but is still rapid enough to present a significant threat to long-term persistence. Accordingly, the long-term persistence of our population is not a trivial consequence of a negligible environmental threat. Our population size, which is order $10^6$ individuals (except when close to extinction), is considerably smaller than is typical for extant molluscs \citep{stanley_1986}, and can be viewed as a conservative lower bound (in any case, $v_j$ only increases logarithmically with $N$ in Eq. \eqref{eq:vmm}). These results provide a rare bridge between micro-evolutionary population genetic models and macro-evolutionary phenomena

The fossil record contains many instances of abundant, widely-distributed species that have suddenly disappeared. This is commonly cited in support of a catastrophic view of extinction \citep{raup_1994}. In contrast, Darwin seems to have regarded sudden disappearance as a fossilization artifact, holding that species typically disappear gradually ``first from one spot, then from another, and finally from the world'' \cite[pp. 317]{darwin_1859}, driven by inter-specific competition \citep{raup_1994}. Thus, there is no need to ``invoke cataclysms to desolate the world'' \cite[pp. 73]{darwin_1859}. These viewpoints share the questionable assumption that sudden disappearance --- assuming it is not an artifact --- indicates a severe, sudden driver of extinction. This clearly need not be true in light of the suddenness of extinction in our gradualist model (Fig.\ \ref{fig:beforeext}), which would appear as a long period of relatively stable abundance followed by sudden disappearance. Sudden disappearance is driven entirely by gradual evolutionary processes, not the one or few extreme environmental changes that characterize catastrophes. In a sense it is still a ``catastrophe'' --- an abnormally large fitness fluctuation --- but this fluctuation reflects poor adaptive performance just as much as environmental pressure. Thus, sudden disappearance alone does little to distinguish between catastrophic or gradual extinction scenarios. The case for a catastrophic interpretation is much stronger if many thriving taxa disappear synchronously (mass extinction), but this excludes much of the fossil record of extinction \citep{raup_1994}, which could therefore plausibly be driven by gradual processes instead.

Long-term evolution in the vicinity of a fitness attractor is a form of Red Queen evolution in which fitness gains are continually thwarted by environmental deterioration, resulting in effectively stagnant mean absolute fitness \citep{van_1973}. As a consequence, persistent populations will have exponentially distributed times to extinction $t_e$, because extinction risk will be essentially independent of population age (ignoring short-term fitness fluctuations). However, if fitness is initially low, say because young populations tend to be colonizers in unfamiliar environments, then the risk of early extinction will be elevated (Fig.\ \ref{fig:extimehist}), and older populations will be less extinction-prone than younger ones. Intriguingly, fossil genera do exhibit reduced extinction risk with age, even after controlling for geographic range and species richness \citep{finnegan_2008}. Our results raise the possibility that population-level evolutionary processes contribute to this pattern (even without major differences in mutation rate or population size), provided that the predicted initial elevation of extinction risk lasts long enough to leave a fossil signature. 

It is interesting to consider the role of genetic load in our model, since different interpretations of load have featured prominently in previous discussions about adaptation rates and extinction risk. In particular, substitutional load arguments directly contributed to the formulation and popularity of neutral theory \citep{kimura_1968}, but their interpretation was controversial. Kimura argued that most substitutions must be neutral because a many-locus version of Haldane's single-locus substitution implies extremely large substitutional loads. However, calculating a cost of selection in this way presumes that the perfect genotype (i.e. with the fittest allele at all loci considered) is present in the population \cite[pp. 78]{ewens_2004}. This effectively conflates the relative substitutional load (proportional to the fitness of the fittest genotype present minus mean fitness; \citealt{crow_1968}) with absolute lag load (proportional to the fitness of the perfect genotype minus mean fitness; \citealt{smith_76}). In our model, the substitutional load is $q_js$, a crucial determinant of the rate of adaptation $v_j$, while the lag load is $\overline{i}s$, a measure of extinction risk. The two are interdependent. In steady state this follows immediately from Eq.\ \eqref{eq:vmm}: for given population parameters, the steady state values of $q_j$ depend on $j$ (and therefore mean fitness $\overline{i}$). Or, looking at it another way, the fitness advantage of new mutants $q_js$ determines $v_j$, which in turn determines the location of the fitness attractor (and if one exists). Interdependence between these relative (substitutional) and absolute (lag) loads is a natural consequence of evolution on an absolute fitness axis driven by relative fitness differences. Substitutional and lag load can therefore be seen as complementary --- but not independent --- aspects of a population's long-term evolutionary status.

Our model is superficially similar to models of mutation load accumulation \citep{lynch_93,kondrashov_1995}, where, instead of environmental change, deleterious mutations gradually erode fitness. However, the effects of accumulating deleterious mutations throughout the population is potentially considerably more complicated, and much weaker, than the population-wide fitness deterioration induced by environmental shifts. For the large asexual populations considered here, provided that the deleterious mutation rate $U_d$ is not very large ($U_d/s\ll 1$), deleterious mutations have little effect on the overall rate of fitness gain regardless of their fitness effect \citep{desai_2007}. If $U_d$ is large enough, a reversible Muller's ``ratchet'' will begin to turn, shifting the entire population one fitness class at time, much like environmental change. However, either $U_d$ would need to be very large or $N$ very small for this mutation-induced deterioration to overpower beneficial mutant establishment and pose an extinction risk \citep{xiaoqian_2011,goyal_2012}.

Some of the results presented here were anticipated by \cite{burger_95}, particularly the pattern shown in Fig. \ref{fig:meantimesmm}b. A major focus of their analysis is what determines the critical rate of environmental change consistent with persistence ($1/T=\max_j v_j$ in our model). Their modeling assumptions are quite different from ours: sex is obligate with free recombination, and population size is small (at most $512$ individuals) and effectively constant (except for after the population crosses the extinction threshold). Consequently, their population has very little linkage disequilibrium, and $N$ is so small that genetic drift and demographic stochasticity are important factors. Our population has high linkage disequilibrium, and $N$ is large enough that stochasticity only plays a role in the establishment of new beneficial mutations. Our approaches can therefore be viewed as complementary, but given the drastic difference in population sizes, it is hard to compare any of their specific predictions to ours.

Probably the biggest limitation of our model is that there is no genetic recombination. Sexual recombination is nearly universal among fossil species and the macroorganisms of interest to conservation biologists. This makes no difference for small populations, which are in the successional regime regardless of recombination. But for large populations, recombination substantially increases the rate of adaptation $v_j$ \citep{neher_2010,weissman_2012}. For relatively simple models of recombination, this increase is fairly well understood \citep{neher_2010,weissman_2012,neher_2013}. Changing $v_j$ does not alter the basic qualitative features of our model, particularly the central role of the fitness attractor, but would affect the quantitative predictions of persistence for given population parameters. 

We have assumed that evolution slows down as the population approaches perfection because the availability of beneficial mutations is limited, represented by the fitness-dependent mutation rate $Uj$ (running out of mutations (RM)). An alternative mechanism for slowing evolution is that beneficial mutations are less effective on fitter genetic backgrounds (diminishing returns epistatis (DR)). Since the relative importance of these alternatives is unresolved \citep{wiser_2013,good_2014}, we checked whether our model is sensitive to the choice of RM versus DR (Supplement C). Even for relatively strong DR, the main effect of using DR instead of RM is causing $v_j$ to have a greater peak value which occurs at lower fitness. This does not alter our conclusions.

\subsection*{Acknowledgements}

This work was financially supported by Wissenschaftskolleg zu Berlin and the National Science Foundation (DEB-1348262). KG is funded by NIH Grant GM084905. We thank Taylor Kessinger for early contributions to the project and Taylor Kessinger, two anonymous reviewers and the associate editor for constructive comments on the manuscript, and Mike Sanderson and Karl Flessa for helpful discussions.

\pagebreak
\nocite{fumagalli_2015}
\bibliographystyle{plainnat}
\bibliography{references}

\pagebreak
\section*{Appendices}

\subsection*{A: Stochastic mutant establishment}
\setcounter{equation}{0}
\renewcommand\theequation{A\arabic{equation}}

Here we summarize how the establishment probability $p_{j-1}$ for a new mutant in fitness class $j-1$ is calculated. 

The mutant lineage's abundance is modeled with a continuous-time birth-death process with time-dependent per-capita birth $\lambda(t)=b(1-N(t)/\kappa)$ and death $\mu(t)=d+(j(t)-1)s$ rates, where the time dependence of $j$ indicates that the initial fitness class of the mutant is $j-1$, but will change to $j$ if the environment deteriorates. This yields \cite[Eq.\ 16]{uecker_2011}
\begin{equation}
p_{j-1}=2 \left[1+\int_{t_m}^\infty (\lambda + \mu) e^{-\int_{t_m}^t (\lambda-\mu) dt'} dt\right]^{-1}, \label{eq:pi}
\end{equation}
where $t_m$ is when the mutant is born. 

Environmental change increases the mortality rate of each fitness class by $s$. Since the population will almost always be in demographic equilibrium  before an environmental change (births $\approx$ deaths; see Appendix C), each fitness class's death rate will exceed its birth rate by $s$ immediately after an environmental change. In particular, the mutant's fitness advantage will be reduced by $s$ in Eq. \eqref{eq:pi}, but will be rapidly restored to $q_js$ as $N$ falls to its new carrying capacity and births balance deaths again. The numerical implementation of Eq. \eqref{eq:pi} in this case is discussed in Supplement A.

For mutants which are undisturbed by environmental change while attempting to establish, we have births $\approx$ deaths (Appendix C), and $\lambda-\mu\approx q_js$. Eq. \eqref{eq:pi} can then be evaluated analytically, yielding $p_{j-1} \approx q_js/(d+j s)$. This is the establishment probability used for our MC approximations. 

\subsection*{B: Markov chain approximation}
\setcounter{equation}{0}
\renewcommand\theequation{B\arabic{equation}}

First, we discuss the MC approximation for the successional regime. The number of adaptive mutant establishments $k_a$ which occur over the time required for fixation of a newly established mutant $t_f$ (see Appendix B) when the population is in fitness class $j$ is Poisson distributed with mean $t_f v_j$, where $v_j$ is given by Eq.\ \eqref{eq:vsuccessional} and $t_f v_j \ll 1$ (Eq.\ \eqref{eq:succ}). Thus, ${\rm Prob}[k_a=0]\approx 1-t_fv_j$, ${\rm Prob}[k_a=1]\approx t_fv_j$ and ${\rm Prob}[k_a>1]\approx 0$. Similarly, $t_f \ll T$ (we are not interested in the case where $T$ is much smaller than $1/v_j$, which implies catastrophically fast environmental deterioration), and the probabilities that $k_e$ mutants establish in $t_f$ generations are ${\rm Prob}[k_e=0]\approx 1-t_f/T$, ${\rm Prob}[k_e=1]\approx v_j/T$ and ${\rm Prob}[k_a>1]\approx 0$. 

The above probabilities for $k_a$ and $k_e$ can be used to define the transition probabilities in an MC with iteration time of $t_f$ and states $0\leq j \leq i_e$. For simplicity, we instead use an MC with an iteration time of one generation, and transition probabilities
\begin{align}
P(j \rightarrow j+1)&= {\rm Prob}[k_e=1] {\rm Prob}[k_a = 0] = 1/T \nonumber \\
P(j \rightarrow j-1)&= {\rm Prob}[k_e=0] {\rm Prob}[k_e = 1]= v_j \nonumber \\
P(j \rightarrow j) &= 1 - 1/T - v_j, \label{eq:transprob}
\end{align}
which behaves essentially identically because fixation takes more than one generation (there are simply more iterations between transitions).

The MC for memoryless adaptation in the multiple mutations regime can be derived in the same way, where the fixation time $t_f$ is replaced by the time required to restore the steady-state distribution following mutant establishment at the nose. Again, this iteration time can be replaced by a single generation iteration time, yielding Eq.\ \eqref{eq:transprob} as before, except that $v_j$ is given by Eq.\ \eqref{eq:vmm} instead of Eq.\ \eqref{eq:vsuccessional}.

In the MC for periodic adaptation in the multiple mutations regime, the iteration time is the mean establishment time $1/v_j$. One adaptive establishment occurs each iteration. The overall transition probabilities are then obtained from a Poisson distribution for the number of environmental change occurrences $k_e$ per iteration, which has rate parameter $1/v_j T$. Thus, $P(j \rightarrow j-1)={\rm Prob}[k_e=0]$, $P(j \rightarrow j)={\rm Prob}[k_e=1]$, and so on. 

For the memoryless MC approximations (in both the successional and multiple mutations regimes), the mean number of iterations $\langle t_e(j) \rangle$ to get from state $j$ to the absorbing extinction state $j=i_{e}$ can be obtained by iterating the chain once to obtain a system of linear equations
\begin{equation}
\langle t_e(j)\rangle=\sum_{i=j-1}^{j+1}  P(j \rightarrow i) \langle t_e(i) \rangle + 1,\qquad t_e(i_{e})=0. \label{eq:meante_linsys}
\end{equation}
For the periodic-adaptation MC, each iteration represents a state-dependent time increment $1/v_j$, and Eq.\ \eqref{eq:meante_linsys} becomes
\begin{equation}
\langle t_e(j) \rangle =\sum_{i=j-1}^{i_{ext}}  P(j \rightarrow i) \langle t_e(i) \rangle  + 1/v_j,\qquad t_e(i_{e})=0. \label{eq:meantereg_linsys}
\end{equation}
Equations \eqref{eq:meante_linsys} and \eqref{eq:meantereg_linsys} can be solved numerically using standard built in routines e.g. numpy.linalg.solve in Python (analytical solution is straightforward, but the resulting solution is cumbersome \citealt[Eq.\ 2.161]{ewens_2004}).

\subsection*{C: Adaptation in the multiple mutations regime}\label{sec:adrate}
\setcounter{equation}{0}
\renewcommand\theequation{C\arabic{equation}}

Here we derive the population's steady-state rate of adaptation $v_j$ and width $q_j$ in the multiple mutations regime. We also introduce the concept of demographic equilibrium, a prerequisite for this derivation.

The multiple mutations regime is contrasted with the successional regime, which is characterized by such long intervals between mutation establishments that the most fit established class $j$ will fix (reach frequency $\approx 1$) well before the next mutation establishes. Suppose that the population is in equilibrium in fitness class $j+1$ when a new mutant lineage establishes in fitness class $j$. Then the new mutant will initially grow exponentially at rate $s$, with starting population size $1/p_j$. The time $t_f$ required for the new mutant to fix then satisfies $e^{s t_f}/p_j \approx K_j$, so that $t_f \approx \ln(K_j s)/s$. For successional behavior, $t_f$ must be much smaller than the time required for the next beneficial mutant to appear, which can be approximated by $1/v_j$ using Eq. \eqref{eq:vsuccessional} (this is a conservative lower bound assuming that the growing mutant lineage already has its fixation abundance $K_j$). Thus, the successional regime occurs when \citep{desai_2007}
\begin{equation}
t_f\approx\ln (K_j s)/s \ll 1/v_j. \label{eq:succ}
\end{equation} 

The multiple mutations regime occurs when mutant lineages establish so frequently (due to high mutation rate or large $N$) that Eq.\ \eqref{eq:succ} is violated. The successional equilibrium $N\approx K_j$ is then never realized, but the population will usually be in a state of approximate demographic equilibrium $N\approx \overline{K}$, where $\overline{K}=\sum_i K_i n_i/N$ is the time-dependent population average of the fitness-class-specific carrying capacities $K_i$ (henceforth, overlines will denote population averages). This demographic equilibrium holds because $N$ changes much faster than $\overline{K}$. With the exception of environmental change events, $\overline{K}$ changes at a rate of 
\begin{equation}
\frac{d\overline{K}}{dt}=-\frac{s \kappa}{b} \frac{d\overline{i}}{dt},
\end{equation}
where $\overline{i}$ obeys Fisher's theorem $d\overline{i}/dt=-s \overline{(i-\overline{i})^2}$ (this version of Fisher's theorem is easily derived from Eq.\ \eqref{eq:dni_bd}; see \citealt[pp. 10]{kimura_1968}). Thus, $d\overline{K}/\kappa dt$ is typically of order $s^2$ between environmental disturbances. By comparison, immediately following an environmental change, $N$ returns to the new value of $\overline{K}$ at exponential rate $s$ ($dN/Ndt=-s$ at the moment after the change), much faster the change in $\overline{K}$ of order $s^2$. Consequently, the per-capita birth rate is approximately $b(1-\overline{K}/\kappa)$ (except in short intervals following environmental change). In the remainder of this Appendix, we assume that this demographic equilibrium holds.

In the multiple mutations regime, mutant establishment follows an inhomogeneous Poisson process driven by mutations in the fittest established class $j$. Over the time interval required for the next mutant to establish, the growth in fitness class $j$ is approximately exponential with rate $(q_j-1)s$ (after the next establishment this growth rate will start to decline appreciably --- see below) and expected starting abundance $1/p_j$ (Eq. \eqref{eq:nu_dist}). Thus, the expected number of mutant lineages that will have established after time $t$ is $Uj b(1-\overline{K}/\kappa) p_{j-1} \int_{0}^{t} e^{(q_j-1)st'}/p_j dt'\approx Uj b(1-\overline{K}/\kappa)\int_{0}^{t} e^{(q_j-1)st'} dt'$ (the approximation uses the fact that $p_{j-1} \approx q_js/(d+j s)$; see Appendix A). Setting this expected number equal $1$, we can solve for the typical time $t$ required for a newly established nose to produce the next mutant lineage that establishes, denoted $t_{est}$. This gives $t_{est}=\ln [(q_j-1)s/Uj b(1-\overline{K}/\kappa)  +1]/(q_j-1)s \approx \ln [s/U j]/(q-1)s$ where $O(1)$ terms inside the logarithm have been ignored (it will become clear below that $q_j$ is $\sim O(1)$). Thus, $t_{est}$ only depends weakly on $j$ over the scale of the population's fitness variation $q_j$ (from Eq.\ \eqref{eq:vmm}, $q_j$ is $\sim O(1)$ for the parameter regime considered here), but varies significantly over the entire fitness domain ($i_{e}\approx (b-d)/s$ is typically $\sim O(10)$ or $\sim O(10^2)$). 

Any given fitness class keeps growing after its initial establishment until it is the most abundant fitness class, with abundance of order $\overline{K}$. During this process, the mean fitness $\overline{i}$ advances, and the fitness class's fitness advantage --- and growth rate --- declines. In steady state, this decline in fitness advantage can be approximated as a sequence of discrete decreases of magnitude $s$ occurring once every $t_{est}$ generations \citep{desai_2007}. Thus, the most recently established mutant, which has a mean initial abundance of $1/p_j$, grows at rate $(q_j-1)s$ for $t_{est}$ generations, then $(q_j-2)s$ for another $t_{est}$ generations, and so on, until it has abundance $\overline{K}$ and no fitness advantage. Thus, $\overline{K}\approx \exp ((q_j-1)s t_{est} + (q_j-2)s t_{est} + \ldots)/p_j$. This implies that the mean fitness advances at a rate of approximately $s^2 q_j(q_j-1)/2\ln (K_j s)$ (again neglecting $O(1)$ terms in the logarithm). In steady state, this must match the rate of advance of the nose $s/t_{est}$; setting them equals gives Eq.\ \eqref{eq:vmm}.

\pagebreak

\section*{Supplement}

\subsection*{A: Numerical implementation of full simulation}

Here we summarize how the full traveling wave model, illustrated in Fig. \ref{fig:travellingwave}, is implemented numerically. 

Eq.\ \eqref{eq:dni_bd} is a system of coupled, nonlinear ODEs describing the dynamics of the bulk of the population. For numerical efficiency, we solve this system over the set of roughly $2q \ll i_{e}$ non-empty abundance classes ($n_i<1$ is regarded as empty), where the set of non-empty classes changes over time and must be updated dynamically. We start by solving Eq.\ \eqref{eq:dni_bd} for these classes from $t=0$ up to the first environmental change at $t=T_1$, where $T_1$ is sampled from an exponential distribution with mean $T$.

Using the resulting solution, we determine the time $t_m$ until the next mutant is produced by the fittest established class $j$. These births follow an inhomogeneous Poisson process with dynamic rate parameter $Uj b(1-N(t)/\kappa) n_j(t)$. To sample $t_m$, we sample a normalized waiting time $\tau$ from an exponential distribution with rate parameter equal to unity, and then find the equivalent waiting time until the next mutation (in generations) for the inhomogeneous Poisson process by solving
\begin{equation}
\tau=\int_{0}^{t_m} Uj b(1-N/\kappa) n_j dt
\end{equation}
for $t_m$. 

If $t_m$ is smaller than $T_1$, we check whether the variant will establish, which occurs with probability $p_{j-1}$ (Appendix A). The numerical evaluation of  Eq.\ \eqref{eq:pi} is computationally expensive, so we use the following approximation scheme:
\begin{enumerate}

\item If the mutant does not arise near an environmental change event, demographic equilibrium implies $p_{j-1}\approx qs/(d+js)$.

\item If an environmental change occurred shortly before $t_m$, the mutant lineage's fitness advantage is reduced by $s$ (Appendix A). We check for this scenario as follows. The lineage's fitness advantage takes $s/[d(\lambda-\mu)/dt|_{t=t_m}]$ generations to change by $s$.  The disturbance from a recent environmental change is important if this timescale is comparable to or shorter than the decay timescale $\sim 1/qs$ of the integral in Eq.\ \eqref{eq:pi}. We then evaluate  Eq.\ \eqref{eq:pi} analytically assuming constant-$N$ as for the case of demographic equilibrium, except that $N$ is averaged over the interval $(t_m,t_m+qs)$. 

\item If the next environmental change is going to occur soon after $t_m$, the value of $p_i$ is sensitive to the timing of the change and must be evaluated numerically. To do this, we solve equation Eq.\ \eqref{eq:dni_bd} in an interval after $T_1$.
\end{enumerate}
The relative error of the resulting approximation for $p_i$ rarely exceeds $10\%$. 

If it is determined that the lineage does establish, at time $t_m$ we remove any fitness classes with abundance $<1$ and add the new fitness class with initial population size $\nu$ sampled from Eq.\ \eqref{eq:nu_dist}. We then repeat the above, starting with solving Eq.\ \eqref{eq:dni_bd} over the interval from $t_m$ to $T_1$.

If $t_m$ exceeded $T_1$, we remove any fitness classes with $n_i<1$ at $T_1$ and repeat the above starting with solving equation Eq.\ \eqref{eq:dni_bd} over the interval from $T_1$ to the next sampled environmental change time $T_2$.

The algorithm terminates if all fitness classes are removed; $t_e$ is defined as the time when $N=1$. 

\pagebreak

\subsection*{B: Duration of observable progression to extinction}

\setcounter{equation}{0}
\renewcommand\theequation{S\arabic{equation}}

Here we analyze the population's behavior in the period immediately preceding extinction using our MC approximation. This analysis supplements the simulation results shown in Fig.\ \ref{fig:beforeext}. 

We run our MC backwards starting at time $t_e$, conditional on extinction occurring at $t_e$. The reverse time process is constructed as follows. Let $Y_0,Y_1,\ldots,Y_{t_e}$ denote the sequence of random variables describing repeated iteration of the reverse-time process i.e. $Y_\tau=X_{t_e-\tau}$ where $X_0, X_1,\ldots,X_{t_e}$ are the random variables for repeated forward-time iteration up to time $t_e$ and $\tau$ measures time before $t_e$. Then, in the absence of conditions on when extinction occurs, the standard expression for backward-time transition probabilities holds:
\begin{equation}
\hat{P}_\tau(i\rightarrow j)={\rm Prob}[Y_{\tau+1}=j|Y_\tau=i]=\frac{p_j(t_e-\tau-1)}{p_i(t_e-\tau)}P(j\rightarrow i), \label{eq:Phatnocon}
\end{equation}
where $p_i(t)={\rm Prob}[X_t=i]$ and $P$ denotes forward-time transition probabilities (Appendix B).

To ensure that extinction occurs at time $t_e$, all three terms in Eq.\ \eqref{eq:Phatnocon} must be made conditional on $t_e$. For $i,j\neq i_e$, $p_i(t_e-\tau)$ becomes
\begin{equation}
{\rm Prob}[X_{t_e-\tau}=i|t_e]=\frac{{\rm Prob}[t_e|X_{t_e-\tau}=i]}{{\rm Prob}[t_e]}p_i(t_e-\tau),
\end{equation}
and similarly for $p_j(t_e-\tau-1)$, whereas $P(j\rightarrow i)$ becomes
\begin{align}
{\rm Prob}[(X_{t_e-\tau}=i|X_{t_e-\tau-1}=j)|t_e]\nonumber\\
=\frac{{\rm Prob}[t_e|X_{t_e-\tau}=i]}{{\rm Prob}[t_e]}P(j\rightarrow i).
\end{align}
Thus, the conditional reverse time transition matrix for $i,j\neq i_e$ is
\begin{align}
\hat{P}_\tau(i\rightarrow j|t_e)&=\frac{{\rm Prob}[t_e|X_{t_e-\tau-1}=j]}{{\rm Prob}[t_e]} \hat{P}_\tau(i\rightarrow j)
\label{eq:Phatcon}
\end{align}

For our problem of reconstructing the pre-extinction behavior of populations which persist for long times, the reverse time transitions Eq.\ \eqref{eq:Phatcon} are independent of $\tau$ to an excellent approximation. Long-term persistence implies that in the period preceding extinction, the forward-time process has had enough time to reach quasi-equilibrium (i.e. the probability of finding the chain in a given state conditional on non-extinction is independent of the initial state --- the initial state has been ``forgotten"). Then $\hat{P}_\tau(i\rightarrow j)$ becomes 
\begin{equation}
\hat{P}(i\rightarrow j)=\frac{p_j(t_e-\tau-1)}{p_i(t_e-\tau)}P(j\rightarrow i) = c\frac{m_j}{m_i}P(j\rightarrow i) \label{eq:qsfirstuse}
\end{equation}
where $m_j$ is the quasi-stationary probability of being in state $j$ (i.e. $m_j=\lim_{t\rightarrow\infty} m_j(t)$ where $m_j(t)={\rm Prob}[X_t=j|{\rm not\, extinct}]$), and $c=1/(1-\sum_{i\neq i_e} m_i P(i\rightarrow i_e))$ is a normalization constant. The quasi-stationary distribution $m_j$ can be obtained by repeated iteration of the forward-time process using the fact that $m_j(t)=p_j(t)/(1-p_{i_e}(t))$.

The term ${\rm Prob}[t_e|X_{t_e-\tau-1}=j]/{\rm Prob}[t_e]$ in Eq.\ \eqref{eq:Phatcon} is more troublesome. It must equal $1$ if $\tau$ is sufficiently large and $j$ is sufficiently distant from $i_e$, since then the condition $X_{t_e-\tau-1}=j$ has no bearing on the risk of extinction at the distant time $\tau$ in the future. To gain some insight into when this breaks down, we can rewrite the term as
\begin{equation}
\frac{f(\tau,j)}{\sum_i f(\tau,i) m_i}
\end{equation}
where $f(\tau,k)={\rm Prob}[t_e,X_{t_e-\tau-1}=j|{\rm not\, extinct}]$. In other words, the ratio deviates from $1$ when $f(\tau,j)$ deviates from its expectation with respect to the quasi-stationary distribution. The greatest potential deviations occur at $j$ values near the extinction threshold $i_e$ where $m_j$ is essentially zero i.e. precisely those states incompatible with long-term persistence. Specifically, if $f(\tau,j)$ is localized at these values, say because $\tau$ is so small that extinction is imminent, the deviation from $1$ may be quite large. But clearly this only applies to $\tau$ values that are tiny compared to the duration of the fluctuating traversal process from the fitness attractor to $i_e$. For essentially all values of $\tau$, the deviation can never be so large as to counteract its multiplication by the corresponding near-zero values of $m_j$ in Eq.\ \eqref{eq:Phatcon}. Accordingly, this term can be set to $1$ to an excellent approximation.

Fig.\ \ref{fig:beforeext} shows the expected behavior of the fittest established class $j$ immediately preceding extinction using the reverse-time transition matrix $\hat{P}(i\rightarrow j)$ to calculate the reverse-time state probability distribution $p_j(t_e-\tau|{\rm not\, extinct})$ (we use the memoryless adaptation approximation for $P(j\rightarrow i)$ to avoid the complication of variable size iteration times). For comparison between MC and simulations, we shift the simulation expectation horizontally so that for both the MC and simulations, $\tau=0$ when $i=i_e$. This accounts for the time it takes for the population to die off after crossing the extinction threshold $i_e$ in the simulations, which is not accounted for in the MC approximation. 

\pagebreak

\subsection*{C: Running out of mutations vs. diminishing returns epistatis}

Here we show that a variant of our MC approximation (Appendix B) which uses a diminishing returns (DR) instead of a running out of mutations (RM) mutation model produces similar long-term behavior. 

To model DR, we replace the fixed fitness increment $s$ with geometric increments $s_i \propto R^{-i}$ ($0<R<1$) \citep{fumagalli_2015}, where $s_i$ is the fitness increment between classes $i$ and $i+1$, and the mortality rate in fitness class $i$ is $d+\sum_{k=0}^{i-1} s_k$. Smaller $R$ represents stronger diminishing returns.

With the change from $s$ to $s_i$, it is no longer possible to keep the fitness effect of environmental deterioration independent of $i$ as it is for RM. Environmental change shifts the population backwards by some integer amount, say $k(i)$, and in general the resulting change in fitness ($s_i+s_{i+1}+\ldots+s_{i+k(i)-1}$) will not be the same for all $i$ no matter how $k(i)$ is chosen. We determine $k(i)$ numerically by minimizing the difference between the resulting fitness change and a ``goal'' environmental change fitness effect $s/T$.

The DR mutation rate $U^*$ is independent of fitness. For comparison with RM, $U^*$ is set equal to the RM beneficial mutation rate $Uj$ averaged over the RM quasistationary distribution $m_j$ (see text after Eq.\ \eqref{eq:qsfirstuse} in Supplement B). 

Fig.\ \ref{fig:meantevsT} compares RM and DR mean extinction times as a function of $T$, making the appropriate changes to Eq.\ \eqref{eq:vmm} for the DR case. DR extinction times tend to be larger (for given $s/T$), because $s_i>s$ when $i$ is closer to extinction than perfection. This has multiple effects, increasing the establishment probability, the size of each fitness jump, and the strength of selection in the bulk of the population. This has a powerful combined effect at low fitness, which outweighs the linear (successional), or sub-linear (multiple mutations) low-fitness benefit of greater mutational availability $Uj$.

When diminishing returns is weak ($R=0.97$), RM and DR are very similar (Fig.\ \ref{fig:DRT62R97}). Fixing $T$ for RM, and bringing the RM and DR fitness attractors into approximate agreement by reducing $T$ in the DR model ($T=151$ compared with $T=180$ for RM), the net rate of fitness increase is similar over most of the fitness domain, and the quasistationary distributions almost coincide. Thus, the difference between these models is primarily a rescaling of the adaptation rate $v_j$.  

For stronger diminishing returns ($R=0.94$), the discrepancy is more substantial (Fig.\ \ref{fig:DRT51R54}). Apart from the fact that a larger change in $T$ is required to bring the fitness attractors together ($T=125$ compared with $T=180$ for RM), the shape of $v_j$ is significantly different over the entire fitness domain, with DR $v_j$ several times larger than RM $v_j$ near extinction. The quasistationary distributions have similar shapes, although in the DR case this corresponds to far fewer fitness classes --- a dense cluster of classes at high fitness is essentially unreachable. This has the effect of ``smoothing'' the transition to persistence (Fig.\ \ref{fig:meantevsT}) in much the same way that increasing $s$ in the RM model does: the fitness attractor becomes less and less relevant the fewer states there are in its vicinity to act as a basin of attraction. However, this also implies that long-term evolution is driven exclusively by large effect mutations, which is probably not realistic. Correcting for this by appropriately increasing $i_e$ for the DR model would again restore the basic qualitative structure of our long-term extinction model, particularly the central role of the fitness attractor. Thus, even fairly strong DR does not substantially alter our predictions. 

\setcounter{figure}{0}
\renewcommand\thefigure{S\arabic{figure}}

\begin{figure}
\centering
\includegraphics[scale=0.5]{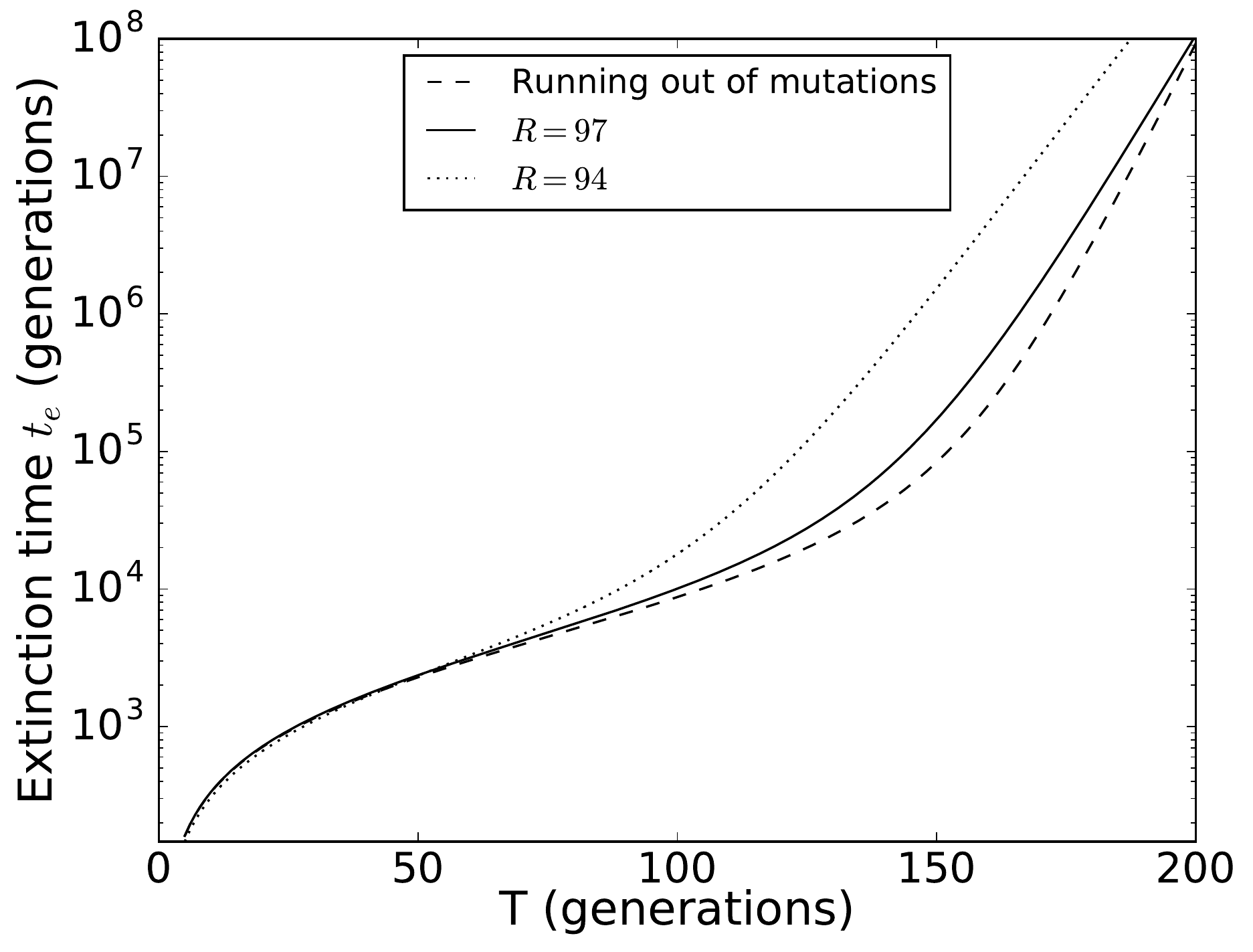} 
\caption{\label{fig:meantevsT} Mean extinction times for weak DR ($R=0.97$) are similar to those for RM. Stronger DR ($R=0.94$) causes a more substantial discrepancy. 
Same parameters as Fig.\ 3.}
\end{figure}

\begin{figure}
\centering
\includegraphics[scale=0.5]{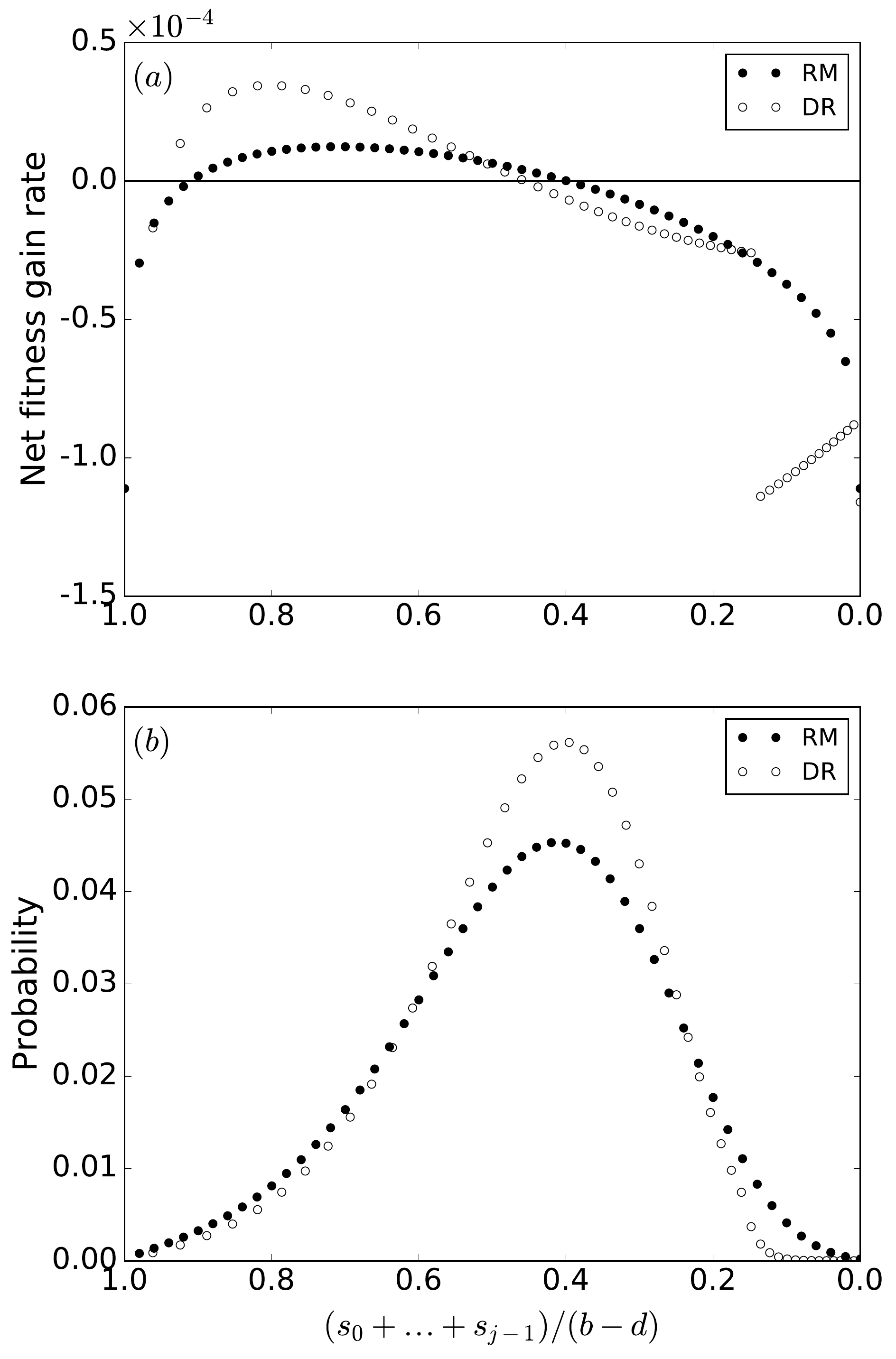} 
\caption{\label{fig:DRT62R97} For weak DR ($R=0.97$), RM and DR models behave similarly aside from greater DR $v_j$. For comparison, we offset the higher DR $v_j$ by adjusting $T$ (DR $T=151$, DM $T=180$). (a) Net rate of fitness increase (i.e. $v_j$ minus environmental deterioration rate) is similar except at high fitness. DR is discontinuous due to the fitting procedure for environmental deterioration fitness jumps. (b) With the $v_j$ offset, quasistationary distributions are essentially the same. Same parameters as Fig.\ \ref{fig:meantevsT} (apart from $T$ offset).}
\end{figure}

\begin{figure}
\centering
\includegraphics[scale=0.5]{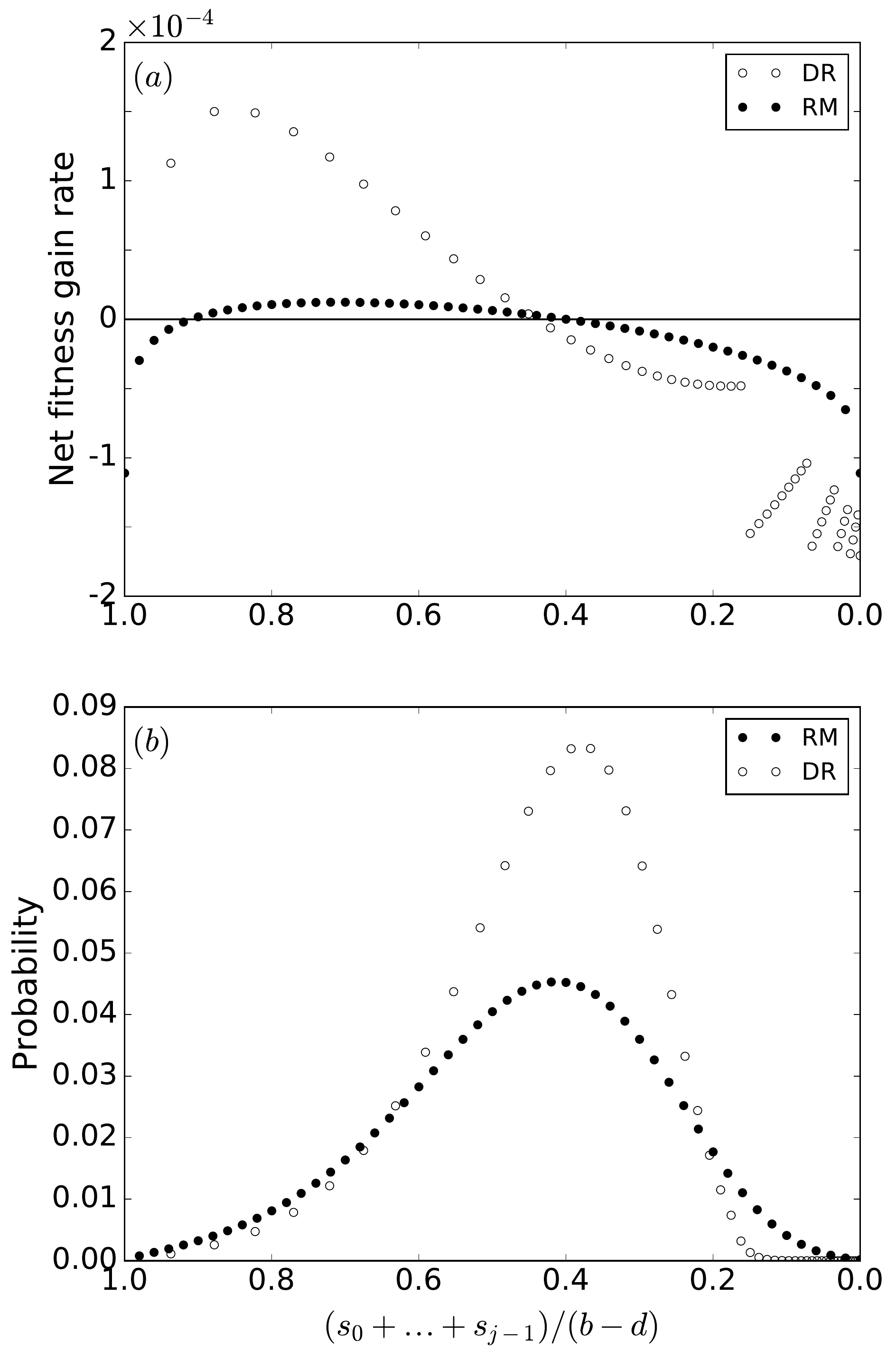} 
\caption{\label{fig:DRT51R54} Same as Fig.\ \ref{fig:DRT62R97} but with stronger diminishing returns ($R=0.94$). Now RM and DR models differ more substantially. As in Fig.\ \ref{fig:DRT62R97}, we offset $v_j$ by adjusting $T$ (DR $T=125$, DM $T=180$). Otherwise same parameters as Fig.\ \ref{fig:DRT62R97}.}
\end{figure}

\end{document}